\documentclass{jaa}
\usepackage[utf8]{inputenc}
\usepackage{amssymb}
\usepackage{amsfonts}
\usepackage{amsmath}
\usepackage[T1]{fontenc}
\usepackage[british]{babel}
\usepackage{latexsym}
\usepackage{times}
\usepackage{etex}
\usepackage{booktabs}
\usepackage{balance}
\usepackage{float}
\usepackage{textcase}
\usepackage{txfonts}
\usepackage{caption}
\usepackage{color}   
\usepackage{epsfig}
\usepackage{graphics}
\usepackage{graphicx}
\usepackage{xmpmulti}
\usepackage{aas_macros} 
\usepackage{astrobib} 
\def\i{\item}

\newcommand{\bed}{\begin{displaymath}}
\newcommand{\eed}{\end{displaymath}}
\newcommand{\bei}{\begin{itemize}}
\newcommand{\eei}{\end{itemize}}
\newcommand{\bef}{\begin{figure}}
\newcommand{\eef}{\end{figure}}
\newcommand{\ben}{\begin{enumerate}}
\newcommand{\een}{\end{enumerate}}
\newcommand{\beq}{\begin{equation}}
\newcommand{\eeq}{\end{equation}}
\newcommand{\ber}{\begin{eqnarray}}
\newcommand{\eer}{\end{eqnarray}}

\newcommand{\pdot}{\mbox{$\dot {\rm P}$}}
\newcommand{\psdot}{\mbox{$\dot{\rm P}_s$}}

\newcounter{attnctr} 

\begin{document}\sloppy

\title{Radio Pulsar Sub-Populations (II) : The Mysterious RRATs} 

\author{Abhishek\textsuperscript{1} and Namrata Malusare\textsuperscript{2}
  and Tanushree N\textsuperscript{3} and Gayathri Hegde\textsuperscript{3}
  and Sushan Konar\textsuperscript{4,*}}
\affilOne{\textsuperscript{1}School of Physics, University of Hyderabad, Hyderabad, 500046, India.\\}
\affilTwo{\textsuperscript{2}Department of Physics, Savitribai Phule Pune University, Pune, 411007, India.\\}
\affilThree{\textsuperscript{3}Department of Physics \& Electronics, Christ (Deemed to be University), Bangalore, 560074, India. \\}
\affilFour{\textsuperscript{4}NCRA-TIFR, Pune, 411007, India.}

\twocolumn[{

\maketitle

\corres{sushan.konar@gmail.com}


\begin{abstract}
Several conjectures have been put forward to explain the RRATs, the newest subclass of neutron stars, and their connections to other radio pulsars. This work discusses these conjectures in the context of the characteristic properties of the RRAT population. Contrary to expectations, it is seen that - a) the RRAT population is statistically un-correlated with the nulling pulsars, and b) the RRAT phenomenon is unlikely to be related to old age or death-line proximity. It is perhaps more likely that the special emission property of RRATs is a signature of them being later evolutionary phases of other types of neutron stars which may have resulted in restructuring of the magnetic fields.
\end{abstract}

}]


\doinum{12.3456/s78910-011-012-3}
\artcitid{\#\#\#\#}
\volnum{000}
\year{0000}
\pgrange{1--}
\setcounter{page}{1}
\lp{1}

\section{Introduction}
Close to $\sim$3500 neutron stars  have been observed and investigated
(in varying detail) since the  serendipitous discovery of PSR B1919+21
in 1967~\cite{hewis68}.  Most  of these neutron stars  are observed as
Radio Pulsars,  characterized by  short spin-periods ($\sim  10^{-3} -
10^2$~s)  and large  inferred surface  magnetic fields  ($\sim 10^8  -
10^{15}$~G). These rotation powered pulsars (RPP) are mostly isolated,
or are members of binaries where active mass transfer is not currently
taking place~\cite{kaspi10,konar13,konar16c,konar17e}.

The  main characteristic  feature  of  a radio  pulsar  (RPSR) is  the
emission  of highly  coherent radiation  (across a  wide range  of the
electromagnetic frequency spectrum) observed as {\em regular periodic}
pulses. However,  the process of  emission is  seen to deviate  from a
regular  pattern  in   a  significant  number  of   RPSRs.   One  such
irregularity is  the phenomenon  of {\em  nulling}, first  detected by
\citeN{backe70} and now seen in $\sim$200 pulsars, which refers to the
abrupt cessation of  pulsed emission for a number of  (can vary from a
few to hundreds or even thousands)  pulse periods. In an earlier work,
we have discussed the population  of Nulling Pulsars (NPSRs) in detail
(\citeNP{konar19e} - Paper-I hereafter).

However,  the extreme  case of  irregular emission  is displayed  by a
group of objects  known as Rotating Radio Transients  (RRAT). In 2006,
eleven  new radio  transient  sources were  discovered  in the  Parkes
Multi-beam Pulsar Survey~\cite{mclau06}.  The  duration of their radio
bursts were $\sim$2-30~ms and the  intervals between the bursts ranged
from minutes to  hours.  None of these were  detectable in periodicity
searches and  appeared only  via single  pulse searches.   The primary
characteristic  of these  transients  appeared to  be sporadic  single
pulse emissions at constant  dispersion measures (DM); with underlying
periodicities suggestive of a neutron star origin.

In this discovery paper, it was conjectured that the RRATs represented
a  distinct and  previously unknown  class of  neutron stars  and were
defined to  be {\em radio pulsars  which can only be  detected through
  single-pulse  searches}. Alternatively,  \citeN{burke10} defined  an
RRAT as a  pulsar which predominantly emits  isolated pulses. However,
many argued that the RRAT behaviour was only an extreme case of pulsar
nulling/intermittency  and required  no separate  identification.  The
justification  for  this  argument  comes   from  the  fact  that  the
log-normal pulse  distributions and power-law frequency  dependence of
the mean  flux are both consistent  with those for the  ordinary RPSRs
(unlike  Magnetars or  giant  pulses  from RPSRs).   This  is a  clear
indication of the RRATs being a subclass of Radio Pulsars whatever may
be the reason for their unusual emission characteristics.

Nevertheless,  with rapidly  increasing number  of RRAT  detections, a
clear-cut  definition   for  these  objects  became   essential.   So,
\citeN{keane11b}  proposed the  following -  ``A RRAT  is a  repeating
radio source, with underlying periodicity, which is more significantly
detectable  via  its single  pulses  than  in periodicity  searches.''
Clearly, this  definition comes with  a caveat. Because, a  pulsar may
appear RRAT-like in a certain observational set-up but not in another.
Notwithstanding the  caveat, all  detections identified as  RRATs till
then~\cite{mclau06,hesse08,shito09,keane10b,burke10,keane11a,burke11}
and    later   observational    campaigns   have    used   the    same
definition. Current public databases, like the `ATNF Pulsar Catalog' or
`The RRATalog',  also use this  same criterion to classify  objects as
RRATs. In  the present work, RRAT  data has primarily been  taken from
these public databases.  Therefore, here too RRAT has the same meaning
as the definition above.

Assuming  these  to represent  a  distinct  sub-population of  neutron
stars, \citeN{mclau06}  estimated that  the RRATs  are expected  to be
significantly more  numerous (by  a factor  of 3  or 4)  than `normal'
pulsars (typically detected through  periodicity searches). Even if we
accept this to  be an over-estimate (given  the uncertainties implicit
in various factors),  the implication is quite serious.   On the basis
of their estimates  of the numbers and birthrates  of Galactic neutron
stars, \citeN{keane08} demonstrated that  treating the normal pulsars,
the  RRATs,  the  XDINS  (X-ray   Dim  Isolated  Neutron  Stars),  the
Magnetars,  and  the  CCOs   (Central  Compact  Objects)  as  distinct
sub-classes  of neutron  stars would  not be  not consistent  with the
observed Galactic supernova rate (required to be greater than or equal
to  total   neutron  star   birthrate).   [See   \citeN{konar16c}  and
  \citeN{konar17e} for  discussions on these  observationally distinct
  sub-populations of neutron stars.]

This  inconsistency  is  suggestive of  connections  (evolutionary  or
otherwise)  between distinct  sub-populations  of  neutron stars.   To
explain  the  `RRAT phenomenon'  (specifically,  the  nature of  their
transient radio emission) a number  of hypotheses have been suggested.
Most of  these hypotheses  have direct bearings  on the  connection of
RRATs  with other  neutron  star sub-populations.  For  many of  these
hypotheses, there exist tentative  observational support. But, as yet,
none of them prove to be definitive.

Bearing these  in mind,  we consider the  characteristics of  the RRAT
population as a whole and examine some of these hypotheses.  The basic
characteristics of  the RRAT population  are discussed in \S2.
In
\S3, we concentrate  upon the question of the connection  of the RRATs
with the  NPSRs, by comparing  these two  populations, in view  of the
results obtained in  Paper-I. In \S4, we consider the  question of the
proximity of the  RRATs and the NPSRs to the  death-line.  A few other
hypotheses regarding  the nature of RRATs  are examined in \S5  and we
summarise our conclusions in \S6.

\section{RRAT Population}
So far 162 RRATs  (listed in Table-[\ref{t_rrat1}-\ref{t_rrat4}]) have
been  found through  archival and  direct pulsar  surveys since  their
discovery~\cite{mclau06}.  Even though RRAT emission has an underlying
periodicity, the  interval between individual pulses  can widely vary,
depending upon the epoch.  Because of this, the spin-period of an RRAT
is  determined  by finding  the  greatest  common denominator  of  the
intervals between  pulses. Clearly, this  method is likely to  yield a
multiple of  the period instead of  the true one, unless  more than a
few pulses  have been  detected in an  observation~\cite{cui17}.  This
problem  is usually  taken  care of  by  detecting sufficiently  large
number of pulses  from a particular RRAT to arrive  at the true period
of the neutron star.

\begin{figure*} 
\vspace{-10.0cm}
\epsfig{file=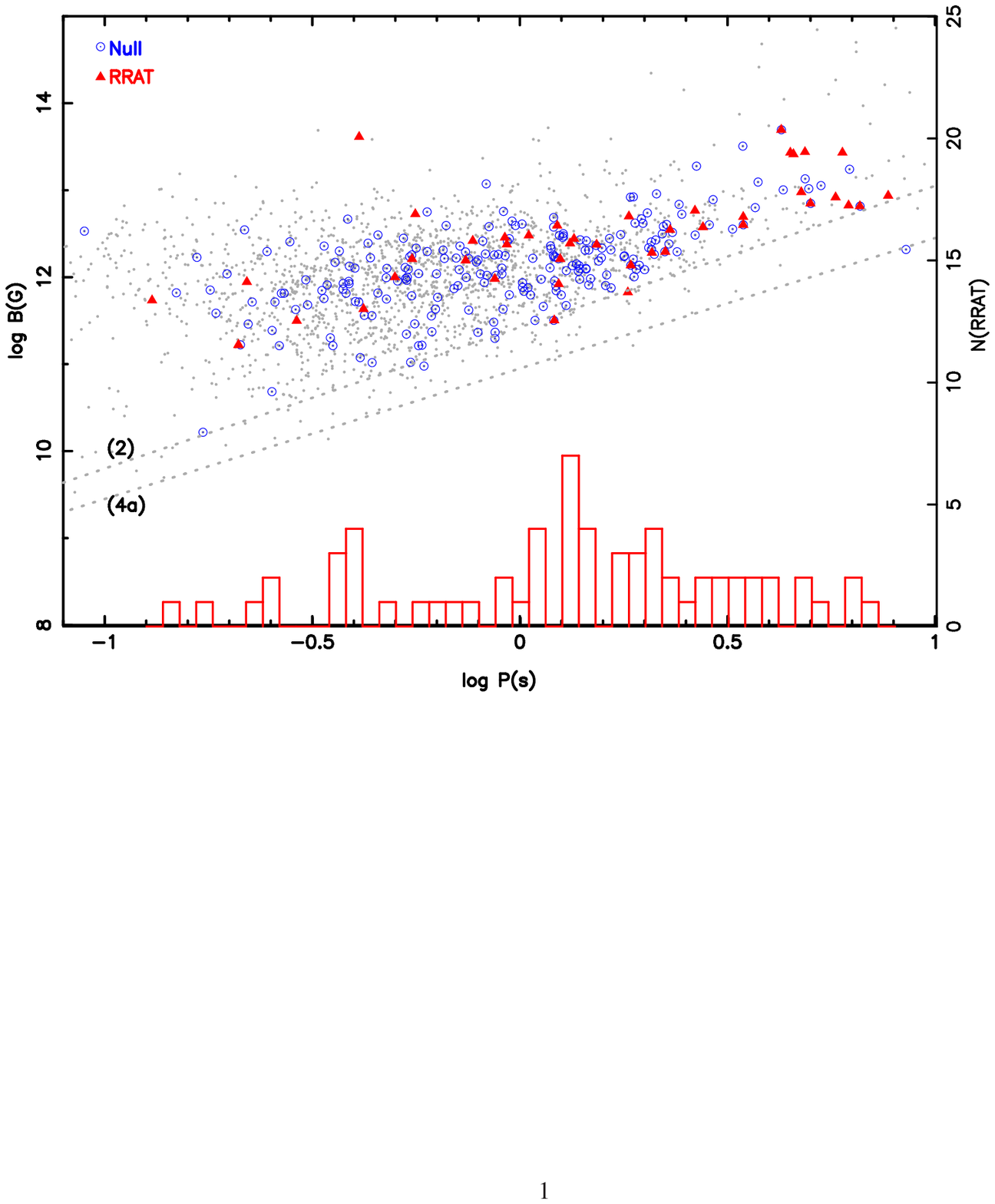,width=600pt}
\vspace{-10.75cm}
\caption{Distribution  of  known  RRATs   (red  triangles)  and  NPSRs
  (circled dots  in blue) along-with  normal RPSRs (grey dots)  in the
  P$_s$--B$_s$ plane.  214  NPSRs (of the 222 known) and  42 RRATs (of
  the 162 known), with estimates of  B$_s$, are shown here.  The P$_s$
  histogram at the  bottom corresponds to another 67  RRATs, for which
  only P$_s$ values  are available. The y-axis on the  right shows the
  number of  objects corresponding  to this RRAT histogram.   The grey
  dashed lines, marked `(2)' and  `(4a)' correspond to two theoretical
  death-lines  (see text  for details).   \\  {\bf Data}  : a)  Normal
  Pulsars - ATNF Pulsar Catalogue, b) RRAT  - Appendix A, \\ c) NPSR -
  Null        Catalogue        (Paper-I        :        \underline{\tt
    http://www.ncra.tifr.res.in/$\sim$sushan/null/null.html}).}
\label{f_bp} 
\end{figure*}

To  date,  the population  of  RRATs  has been seen to have the
following  ranges for  the  spin-period (P$_s$),  the dipolar  surface
magnetic field (B$_s$), and the dispersion measure (DM) --
\vspace{-0.25cm}
\bei
   \i P$_S$ :  41.5~ms -- 7.7~s ;
   \i B$_s$ : 1.67 $\times 10^{11}$ -- 4.96 $\times 10^{13}$~G;
   \i DM : 4.0 -- 786.0~pc.cm$^{-3}$.
\eei
\vspace{-0.25cm}
Understandably, timing  measurements are difficult and  values for the
spin-period or the period derivative (hence, estimates for the dipolar
surface magnetic field) are not available  for all of the known RRATs.
In Fig.[\ref{f_bp}], the RRATs (for  which both P$_s$ and B$_s$ values
are  available)  have been  shown,  along  with  the NPSRs  and  other
non-nulling RPSRs, in the P$_s$-B$_s$ plot. A P$_s$ histogram, for the
RRATs that do not have known values  of B$_s$, is also provided at the
bottom of  Fig.[\ref{f_bp}] to  indicate the  general trend  for P$_s$
values. It is seen that the spin-period and the surface magnetic field
values of RRATs  are skewed towards the higher side  compared to those
for the normal RPSR population  (millisecond pulsars excluded).  It is
also noted that  many RRATs occupy the same region  of the P$_s$-B$_s$
plane where some Magnetars have been found~\cite{cui17}.

Now, the basic requirement for pulsar  emission is a copious amount of
pair production  in the  magnetosphere of  the neutron  star.  Pulsars
`switch  off'  when   conditions  for  pair  production   fail  to  be
met. \citeN{ritch76} was the first to  define a cut-off line (known as
the {\em  death-line} now) for pulsar  emission in the context  of the
NPSRs.   It was  conjectured  that the  NPSRs  experience null  simply
because of  their proximity  to the death-line.   Over the  years, the
theory  of  {\em death-line}s  has  been  investigated in  detail.  In
Paper-I we have discussed the  significance of a number of death-lines
that have been  proposed, in the light of the  current pulsar data. It
is noticed  that two of  the death-lines developed  by \citeN{chen93a}
are of importance for the RPSR population as a whole, and the NPSRs in
particular.

Both of these  death-lines are obtained assuming  the pair productions
($\gamma +  B \rightarrow e^- +  e^+$, $\gamma$ - photon,  $e^{-/+}$ -
electron/positron, B - magnetic field) to occur predominantly near the
polar cap of the neutron  star~\cite{ruder75}. It is also assumed that
a) the  surface field is dipolar,  and b) the radius  of curvature for
the  magnetic field  is  approximately equal  to  the stellar  radius.
These death-lines, numbered as {\bf [2]} and {\bf [4a]} in Paper-I are
as follows -
\ber
\mbox{\bf [2]} &\mbox{\bf :}& 4 \log B_s - 6.5 \log P_s = 45.7 \,, \\
\label{eq-dl2}
\mbox{\bf [4a]} &\mbox{\bf :}& 4 \log B_s - 6 \log P_s = 43.8 \,;
\label{eq-dl4a}
\eer
where P$_s$ is in seconds and  B$_s$ is in Gauss. Death-line {\bf [2]}
corresponds to the case of very  curved field lines whereas {\bf [4a]}
is obtained for extremely twisted field lines. It is clearly seen from
Fig.[\ref{f_bp}]  that  the  region beyond  death-line~{\bf  [4a]}  is
almost empty.  However, the  RRATs and the  NPSRs are  almost entirely
bounded below  by the  death-line~{\bf[2]}, suggesting  similarities /
connections between these two populations.

\section{RRAT-NPSR Connection}
Because of their  intermittent nature it has been natural  to look for
similarities   and/or   connections   between  the   RRATs   and   the
NPSRs. \citeN{burke10}  considered this issue in  detail and concluded
that RRATs are likely to be  `extreme' cases of NPSRs. They found that
PSR~J0941-39 switches between a RRAT-like  and an NPSR-like mode, i.e.
sometimes appearing  with a sporadic  RRAT-like emission and  at other
times emitting  as a bright  regular nulling pulsar. Noting  that this
object may represent a direct  link between ordinary pulsars and RRATs
they suggested  that RRATs could  be an evolutionary phase  of pulsars
with a high  nulling fraction (NF -  fraction of time a  pulsar is not
seen in  emission) or nulling pulsars  that `switch on' for  less than
the duration of a pulse period.

An object of interest in this context is PSR~J1107-5907, known to have
many different modes of emission -  a strong mode with a broad profile
with nulls,  a weak  mode with  a narrow  profile that  has occasional
bursts of  up to  a few  clearly detectable  pulses at  a time,  and a
low-level underlying  emission. It  has been  argued that  this source
would look like  an RRAT for most  of the time, if placed  at a larger
distance~\cite{young14}. A very similar  conjecture has also been made
about another  NPSR (B0656+14/J0659+1414) that it  would have appeared
RRAT-like if located at a greater distance~\cite{welte06}.  J1107-5907
also exhibits  different NFs in  different emission modes,  similar to
that observed  in B0826-34  (J0828-3417) and  J0941-39, both  of which
appear to switch between an RRAT-like and a more typical ‘pulsar-like’
phase~\cite{burke10,burke12,esamd12}.

Then  again, the  weak  emission  mode of  B0826-34  is  liable to  be
confused with  nulling phases if the  signal is not integrated  over a
sufficiently long  interval of  time.  This has  been suggested  to be
indicative of an evolutionary  progression towards the death-line; and
that all  (or most) pulsars  likely start off as  continuous emitters,
gradually begin  to null and then  increase their NF to  become RRATs,
ultimately crossing  the death-line to  end the active  radio emission
phase~\cite{burke10}.  In fact, both J1107-5907 and B0826-34 are quite
close to the death line.  However,  there also exist a large number of
NPSRs that  are close to the  death-line but are not  known to exhibit
any RRAT-like behaviour.   Similarly, quite a few RRATs  are found far
away from  the death-line (see Fig.\ref{f_bp}).  Therefore, death-line
proximity  does  not immediately  imply  a  certain nature  of  pulsar
emission. We discuss this issue in detail in \S3.

\begin{table}[h]
\centering
\begin{tabular}{rllll} \toprule
     & P$_s$ & B$_s$       & $\tau_c$    & DM         \\ 
     & s     & 10$^{12}$~G & 10$^7$~yr & pc.cm$^{-3}$ \\ \midrule
RRAT & 2.07  & 7.18       & 1.59      & 105.91      \\ 
N    & 109   & 44         & 44        & 162         \\ \midrule
NPSR & 1.26  & 2.63       & 3.99      & 134.78      \\
N    & 222   & 215        & 215       & 222         \\ \midrule
ALL  & 1.53  & 3.40       & 3.58      & 122.60      \\
N    & 331   & 259        & 259       & 384         \\  \bottomrule
\end{tabular}
\caption{Average values of P$_s$, B$_s$, $\tau_c$ and DM for the RRAT,
  the NPSR and the combined populations. The rows  marked `N' indicate
  the  number  of  objects  for  which  values  of  the  corresponding
  parameter are  available and have  therefore been used  to calculate
  the average. NPSRs also include the intermittent pulsars. {\bf Data}
  : a) RRAT - Appendix A, b) NPSR - Null Catalogue.}
\label{t_stat-I}
\end{table}

Even though  the picture is not   yet clear, it is  evident that there
exist  certain connections  between the  RRATs and  the NPSRs.  Clear
nulling  segments  were  also   exhibited  recently  by  another  RRAT
(J1913+1330) in  a FAST  observation~\cite{lu19}. At this  point, some
other objects have also been observed to exhibit such RRAT-NPSR `dual'
nature  (see  Table-[\ref{t_ORAT}]).   To  understand  this  connection
better, we  look at the  statistical nature of  the RRAT and  the NPSR
populations below.

Table-[\ref{t_stat-I}]  summarises  the   average  values  of  relevant
physical quantities of the RRAT and the NPSR populations. Clearly, the
RRATs  have larger  average spin-periods,  and larger  average surface
magnetic  fields, as  has already  been remarked  upon.  Consequently,
NPSRs have larger characteristic ages ($\tau_c$) compared to the RRATs
(B$_s$$\sim$$\sqrt{{\rm        P}_s\psdot}$,       $\tau_c$$\sim$${\rm
  P}_s/\psdot$).  We must remember that the estimate of characteristic
age is  crucially dependent on  the assumption of a  constant magnetic
field, which  may or may not  be applicable to RRATs.   However, RRATs
clearly  have  lower   average  DM,  indicating  that   they  are  not
necessarily located farther away compared to the NPSRs. Therefore, the
conjecture  that  a distant  NPSR  might  appear  to be  RRAT-like  is
unlikely to explain the majority of  objects, even if such a situation
is realised for a small number of the RRATs.

\bef
\epsfig{file=rrat_phst.ps,width=150pt,angle=-90}
\caption{P$_s$ distribution of  the RRATs (red), the  NPSRs (blue) and
  the combined (NPSR + RRAT) population (black broken line). Number of
  objects  (data  points) available  for  each  distribution is  shown
  within brackets.}
\label{f_Phst} 
\eef

Fig.\ref{f_Phst}--\ref{f_Dhst}  show  the distribution  histograms  of
these  parameters.    In  order   to  quantify   the  nature   of  the
distributions,  we perform  1-sample Kolmogorov-Smirnov~\cite{mises64}
tests  on   each  set  of   data.   The  results  are   summarised  in
Table-[\ref{t_stat-II}].    It  is   seen  that   P$_{\rm  KS}$,   the
probability of  error for  rejecting the hypothesis  that the  data is
normally  distributed,   is  very  high  for   $\log({\rm  P}_s)$  and
$\log({\rm B}_s)$ but  tiny for DM for both RRAT  and NPSR populations
(P$_s$  in units  of second  and B$_s$  in units  of Gauss).  Clearly,
$\log({\rm P}_s)$  and $\log({\rm B}_s)$  are expected to  be normally
distributed.  In  other  words,   P$_s$  and  B$_s$  are  log-normally
distributed for  both the populations, though  with somewhat different
degree   of    normality.    We   also   perform    the   2-population
Kolmogorov-Smirnov tests on  the parameter values to  compare the RRAT
and the NPSR populations.  The bottom panel of Table-[\ref{t_stat-II}]
show the results.  It is obvious that the statistical distributions of
P$_s$, B$_s$ for RRATs and NPSRs are completely dissimilar.

The combined  distributions, for  both of  these parameters,  are also
log-normally  distributed with  a degree  of normality  in-between the
RRAT and the  NPSR distributions.  This could be due  to the fact that
the size  of the NPSR population  is much bigger (twice  for P$_s$ and
five  times  for  B$_s$  values)  than the  RRAT  population  and  the
behaviour of the combined population  is basically dictated by that of
the NPSRs.   It might be tempting  to conclude that the  NPSRs and the
RRATs  come  from  two  different  segments  of  the  same  underlying
population which  is Gaussian in  nature. But  the fact that  both the
RRAT  and the  NPSR parameters  also  tend to  be separately  Gaussian
contradicts such a conclusion.

As expected, the nature of the DM distribution is very different. Given
that  the dispersion  measure  can be  considered to  be  a proxy  for
distance (with  certain caveats) a  normal distribution is  not really
expected.   Because  every  detection   is  limited  by  the  inherent
sensitivity  of  the particular  observation  and  nearer sources  are
expected  to  be  detected  with  a higher  probability.   But  it  is
surprising to note that the DM distributions of the NPSRs and the RRATs
are totally  dissimilar (P$_{\rm KS}$(R-N)  $< 10^{-2}$). As  both the
RRATs and the  NPSRs are being detected  by same/similar observational
instruments, we can interpret this result in two ways - either, a) the
single pulse searches  (detecting RRATs) are more sensitive  at low DM
compared to  the periodicity searches (detecting  regular pulsars), or
b) for any  given detection sensitivity only RRATs  located at shorter
distances  are  detected.  The  second possibility,  suggestive  of  a
scenario  where the  `RRAT  phenomenon' is  observed  only for  nearby
neutron  stars,  is  problematic  as  it  exacerbates  the  `birthrate
problem' discussed earlier.

In summary, the current populations of the RRATs and the NPSRs can not
be said to belong to the same sub-class of RPSRs. Therefore, while the
observations  clearly   suggest  that   they  have   certain  inherent
connections,  it would  perhaps  not  be accurate  to  treat the  RRAT
emission simply as an extreme form of nulling behaviour.

\bef
\epsfig{file=rrat_bhst.ps,width=150pt,angle=-90}
\caption{B$_s$  distribution of  the RRATs  (red), the  NPSRs and  the
  combined distribution (black broken line). }
\label{f_Bhst} 
\eef
\bef
\epsfig{file=rrat_dhst.ps,width=150pt,angle=-90}
\caption{DM  distribution  of  the  RRATs (red),  the  NPSRs  and  the
  combined distribution (black broken line). }
\label{f_Dhst} 
\eef
\begin{table}[h]
\centering
\begin{tabular}{ r l l l} \toprule
  & log(P$_s$/s) & log(B$_s$/G) & DM          \\ \midrule
\multicolumn{4}{l}{RRAT} \\
 P$_{\rm KS}$(N) & 0.36 & 0.93 & $< 10^{-5}$ \\ 
 D$_{\rm KS}$(N) & 0.08 & 0.07 & 0.23 \\ 
 N     & 109    & 44     & 162    \\ \midrule
\multicolumn{4}{l}{NPSR} \\
 P$_{\rm KS}$(N) & 0.70 & 0.55 & $< 10^{-5}$ \\
 D$_{\rm KS}$(N) & 0.05 & 0.05 & 0.17 \\
 N     & 222    & 215    & 222    \\ \midrule
 \multicolumn{4}{l}{ALL} \\
 P$_{\rm KS}$(N) & 0.47 & 0.63 & $< 10^{-5}$ \\
 D$_{\rm KS}$(N) & 0.05 & 0.05 & 0.19 \\
 N     & 331    & 259    & 384       \\  \midrule
\midrule
\multicolumn{4}{l}{KS2 : RRAT vs. NPSR} \\ 
 P$_{\rm KS}$(R-N) & $< 10^{-3}$ & $< 10^{-3}$ & $\sim 10^{-2}$ \\
 D$_{\rm KS}$(R-N)  & 0.25 & 0.33 & 0.17 \\
 N  & 109,222    & 44,215 & 162,222    \\  \hline
\end{tabular}
\caption{{\bf Kolmogorov-Smirnov test statistics :} P$_{\rm KS}$(N) is
  the probability of error for  rejecting the null hypothesis that the
  sample is normally distributed, and D$_{\rm KS}$(N) is the magnitude
  of the  difference between  the sample  distribution and  the normal
  distribution.    The   tests   are   performed   on   $\log$(P$_s$),
  $\log$(B$_s$) and DM. P$_{\rm KS}$(R-N) \& D$_{\rm KS}$(R-N) are the
  corresponding quantities for the null hypothesis that both the RRATs
  and the NPSRs  are drawn from the same  underlying population.  {\bf
    Data} : a) RRAT - Appendix A, b) NPSR - Null Catalogue.}
\label{t_stat-II}
\end{table}
\section{Death-Line Proximity}
It  has been  suggested that  the RRATs  could be  RPSRs close  to the
pulsar death line  in the  P$_s$-B$_s$ plane.  These  systems might be
emitting weak, continuous  radio pulses, which have  not been detected
yet, in addition to the observed short radio bursts~\cite{welte06}.

To check  if the RRAT  phenomenon is connected  to a proximity  to the
death-line, we define a proximity parameter, $\epsilon$, given by -
\beq 
\epsilon = \frac{\tau_D - \tau_c}{\tau_D} \,,
\label{eq_eps}
\eeq
where, $\tau_c$ is the current  (characteristic) age of a pulsar given
by P$_s$/2\psdot. This definition of $\tau_c$ inherently assumes that -
a) the pulsar has P$_s$ $\sim$0 at birth, and b) B$_s$ does not evolve
significantly over  the active  lifetime of  a pulsar.   This constant
surface  magnetic field  of a  pulsar is  obtained using  the measured
values of P$_s$ and \psdot~from the relation -
\beq
B_s = 3.2 \times 10^{19} \sqrt{{\rm P}_s \dot{\rm P}}\, \, \mbox{G} \,,
\eeq
where P$_s$ is in seconds and \psdot~is in $ss^{-1}$, assuming a purely
dipolar field.

Under the same  assumptions mentioned above $\tau_D$ is  defined to be
the total time taken by a pulsar to reach the death line, such that
\beq
\tau_D = \frac{{\rm P}_D}{2 \dot{{\rm P}}_D} \,,
\eeq
where  P$_D$ is  the spin-period  at death  line and  \pdot$_D$~is the
period  derivative  at  the  death-line. Clearly,  the  definition  of
$\tau_D$ depends  on a particular  choice of the death-line.   In this
work, we  adopt death-line {\bf [2]},  given by Eq.(1). This  gives us
the following expressions for P$_D$ and \pdot$_D$ -
%
\ber
\log {\rm P}_D = 0.62 \log {\rm B}_s - 7.03 \,, \\
\dot{\rm P}_D = 9.76 \times 10^{-39} {\rm P}_D^{-1} \, {\rm B}^2_s \,;
\eer
where P$_D$ is in seconds, \pdot$_D$~is  in $ss^{-1}$, and B$_s$ is in
G.  The  distribution histograms  of  P$_D$  and $\epsilon$,  for  the
corresponding   populations  are   shown   in  Fig.\ref{f_PDhst-R}   -
\ref{f_ehst}.

\bef
\epsfig{file=rrat_pdhst.ps,width=150pt,angle=-90}
\caption{Distribution of P$_s$~(red) for RRATs with available B$_s$ values, marked as RRAT-B. The corresponding distribution of P$_D$ (blue) is marked as RRAT-B@Death. The number of objects for RRAT-B and RRAT-B@Death are different because two RRATs from the first group lie below death-line {\bf [2]}, used for calculating P$_D$. The broken black line represents the P$_s$ distribution of all 108 RRATs for which spin-period has been measured (irrespective of the availability of B$_s$ value). }
\label{f_PDhst-R} 
\eef
\bef
\epsfig{file=null_pdhst.ps,width=175pt,angle=-90}
\caption{Distribution of P$_s$~(blue) for NPSRs with available B$_s$ values, marked as Null-B. The corresponding distribution of P$_D$ (magenta) is marked as Null-B@Death. The number of objects for Null-B and Null-B@Death are different because four pulsars from the first group lie below death-line {\bf [2]}, used for calculating P$_D$.}
\label{f_PDhst-N} 
\eef
\bef
\epsfig{file=rrat_ehst.ps,width=175pt,angle=-90}
\caption{Distribution of $\epsilon$,  as defined by Eq.(\ref{eq_eps}),
  for the RRATs and the NPSRs.  In both the cases, objects included in
  the distributions  are those for  which B$_s$ values are  known, and
  their  current  locations,  in  the  P$_s$-B$_s$  plane,  are  above
  death-line {\bf [2]}. }
\label{f_ehst} 
\eef

It  is to  be noted  that P$_D$  or $\tau_D$  and hence  the proximity
parameter, $\epsilon$, can only be  calculated where the current value
of \psdot~has  been measured.  This  is why  such a calculation  can be
done only for a  small number of RRATs, even though  it is possible to
obtain $\epsilon$ for almost the entire population of NPSRs. Moreover,
for any  given choice of death-line,  there exist a few  objects which
fall on the right of the line (in the P$_s$-B$_s$) plane. This happens
because  every   theoretical  death-line  makes   certain  simplifying
assumptions which  may or may  not be  applicable for a  given pulsar.
So, these 'beyond death-line' pulsars have also been excluded from the
$\epsilon$ calculation.

\begin{table} 
\centering
\begin{tabular}{rllll} \toprule
  & P$_D$ & P$_{\rm KS}$(N) & D$_{\rm KS}$(N) & $\epsilon$ \\ 
RRAT (42)
  & 6.44s & 0.93         & 0.08          & 0.96      \\ 
NPSR (211)
  & 3.52s & 0.57         & 0.05          & 0.96      \\ 
ALL (253)
  & 4.01s & 0.61         & 0.05          & 0.96      \\ \midrule
\midrule
KS2(R-N)  & P$_D$        & P$_D$         & $\epsilon$ & $\epsilon$ \\
  & P$_{\rm KS}$ & D$_{\rm KS}$ & P$_{\rm KS}$ & D$_{\rm KS}$ \\ 
  & $< 10^{-3}$   & 0.36 & 0.90 & 0.09 \\ \bottomrule 
\end{tabular}
\caption{The average values of P$_D$  and $\epsilon$ (see text for the
  definitions)   for   the   RRAT,   the   NPSR   and   the   combined
  populations. Kolmogorov-Smirnov test statistics, P$_{\rm KS}$(N) and
  D$_{\rm  KS}$(N), to  check whether  P$_D$ is  normally distributed.
  (This  test is  nor  performed on  $\epsilon$ as  it  is clearly  an
  asymmetric,  skewed  distribution.)   P$_{\rm KS}$(R-N)  \&  D$_{\rm
    KS}$(R-N) are the corresponding quantities for the null hypothesis
  that both the RRATs and the NPSRs are drawn from the same underlying
  population.   {\bf Data}  : a)  RRAT -  Appendix A,  b) NPSR  - NULL
  Catalogue. }
\label{t_stat-II}
\end{table}
The   average   values   of   P$_D$   and   \pdot$_D$~are   shown   in
Table-[\ref{t_stat-II}].   For   the  NPSRs,   it  can  be   seen  from
Fig.\ref{f_PDhst-N}   that  P$_D$   has   a   very  clear   log-normal
distribution with the peak at $\sim2.5 s$. Though it is not as obvious
for  the RRATs,  the  P$_D$ distribution  is  again nearly  log-normal
albeit with a  higher peak value of $\sim5  s$. Surprisingly, contrary
to expectation,  the proximity  parameter for both  the RRATs  and the
NPSRs  have  rather  similar  behaviour.  Even  though  close  to  the
death-line,  both types  of  objects have  larger  fractions of  their
life-time to  go through  yet. This likely  rules out  the possibility
that nulling or  RRAT behaviour can be treated simply  as an `old age'
characteristic appearing in a  particular evolutionary phase. However,
as  mentioned  above,  such  a conclusion  crucially  depends  on  the
assumption of constant surface magnetic fields which may or may not be
true for the objects under consideration.

Therefore, it is important  to consider other evolutionary connections
for  the RRATs.   In  the  next section  we  discuss  a few  important
conjectures of that nature. 

\section{Other Connections}
\begin{table} 
\centering
\begin{tabular}{llrrl} \toprule
RRAT       & P$_{\rm s}$ & B$_{\rm s}$  & DM          & Other \\
J-Name     & s         & 10$^{12}$~G & pc.cm$^{-3}$ &      \\ \midrule
J0828-3417 & 1.85      & 1.37       & 52.20       & RPSR \\
J0941-39   & 0.59      & ----       & 78.20       & NPSR \\
J1119-6127 & 0.41      & 41.00      & 704.80      & RPSR \\
           &           &            &             & Magnetar \\
J1647-3607 & 0.21      & 0.17       & 224.00      & NPSR \\
J1819-1458 & 4.26      & 49.60      & 196.00      & Magnetar \\
J1840-1419 & 6.60      & 6.55       & 19.40       & NPSR \\
J1854-1557 & 3.45      & 4.00       & 150.00      & NPSR \\ 
J1913+1330 & 0.92      & 2.86       & 175.64      & NPSR \\ 
J2033+0042 & 5.01      & 7.05       & 37.84       & NPSR \\ \bottomrule
\end{tabular}
\caption{RRATs  that  also occasionally  show  up  as other  kinds of
  neutron stars (indicated in the last column). Here, RPSR means radio
  pulsars not known to show any nulling behaviour. }
  \label{t_ORAT}
\end{table}

Initially, the RRAT phenomenon was considered to be a manifestation of
certain selection bias  and / or the sensitivity of  a telescope.  For
example,   they  were   thought   to  be   giant   pulses  from   weak
pulsars~\cite{knigh06};  or  to  be  pulsars,  with  special  emission
behaviour, being far away (as discussed  in \S3). But there appears to
be far more to the RRATs than simple observational effects.

\begin{figure*} 
\epsfig{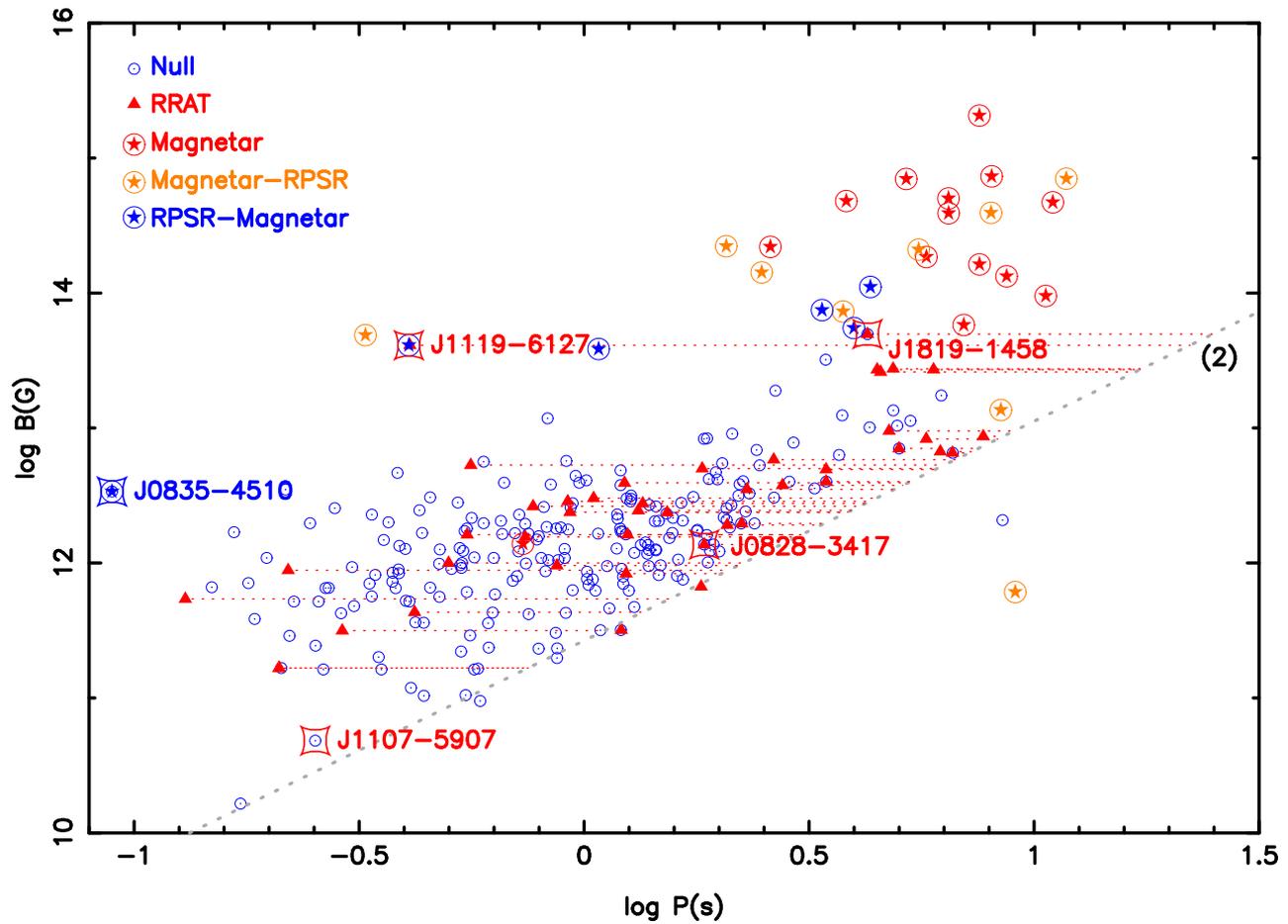}
\caption{Distribution of RRATs (red  triangles), NPSRs (circled dot in
  blue) and  Magnetars (circled stars  in orange) in  the P$_s$--B$_s$
  plane.  Green  circled stars  represent either a Magnetars exhibiting
  RPSR  characteristics or an RPSRs exhibiting Magnetar characteristics.
  The  red  dashed-lines  are  evolutionary tracks  of  RRATs  to  the
  theoretical  death-line  `(2)'  (grey  dashed-line),  with  constant
  B$_s$.  See  text for the  significance of the named  pulsars.  {\bf
    Data  :} a)  RRAT  - Appendix  A,  b) NPSR  -  NULL Catalogue,  c)
  Magnetar       -        Magnetar       Catalog       (\underline{\tt
    http://www.physics.mcgill.ca/~pulsar/magnetar/main.html}).}
\label{f_bp-death} 
\end{figure*}

Besides the NPSRs,  regular non-nulling RPSRs have  also been observed
to behave like RRATs, like J0828-3417 or J1119-6127.  While J0828-3417
is an  ordinary pulsar with  a low  DM, J1119-6127 is  a high-magnetic
field   pulsar    with   a    rather   high    value   of    DM   (see
Table-[\ref{t_ORAT}]). In fact, J1119-6127 has been observed to exhibit
different types of radio behaviour at different epochs, with RRAT-like
events   typically    preceded   by   large    spin-period   glitches.
\citeN{welte11}  argued that  the  glitches could  be responsible  for
reconfiguration of  the magnetic  field giving  rise to  such changed
emission behaviour, and that this  likely indicates the existence of a
group of neutron  stars that become visible for a  brief while only in
the immediate aftermath of glitch activity.

It has also  been suggested that RRATs likely  have evolutionary links
with   the  Magnetars~\cite{mclau09}   or  the   XDINS~\cite{popov06}.
Observation  of  RRATs in  X-ray  would  help confirming  such  links.
Unfortunately,   only   two   RRATs,   J1819-1458~\cite{mclau07}   and
J1119-6127~\cite{archi17},  have been  observed in  X-rays with  upper
limits  for  the X-ray  luminosity  estimated  for another  two  RRATs
(J0847-4316, J1846-0257).   Most probably,  the main reason  for X-ray
non-detection   of    RRATs   is    the   uncertainties    in   source
positions~\cite{kapla09b}.  The post-glitch  recovery of the frequency
derivatives  (decrease in  the average  spin-down rate  instead of  an
increase), as well as the X-ray outbursts of J1819-1458 and J1119-6127
have         been        observed         to        be         clearly
Magnetar-like~\cite{lyne09,rea10,archi17,bhatt18}.  Given  the unusual
change in \psdot, it has been suggested that J1819-1458 is actually
transitioning  from being  a Magnetar  to an  RRAT~\cite{lyne09}.  The
view  that the  RRATs could  be an  evolutionary stage  rather than  a
separate  class  of  neutron  stars is  indeed  strengthened  by  such
observations.

On  the  other  hand,  \citeN{genca18}   has  argued  for  a  case  of
accretion-induced revival  of a  pulsar, that  has evolved  beyond the
death-line, in their  fallback-disc model to explain  the behaviour of
J1819-1458.   This model,  while  reproducing the  observed values  of
P$_s$,  \psdot~and  the  X-ray  luminosity, obtains  a  dipolar  field
strength of $\sim 5 \times 10^{11}$~G at the polar cap.  This value of
the magnetic field  combined with the measured P$_s$  implies a `dead'
pulsar, located beyond  the death-line in the  P$_s$--B$_s$ plane.  If
this indeed  is the case,  then only  accretion can explain  the X-ray
activity.  It is therefore argued  that J1819-1458 is evolving towards
becoming an XDINS, reinforcing an earlier suggestion of the connection
between  RRATs with  such objects.   One  needs to  remember that  the
magnetic field obtained  in this model is much smaller  than the field
inferred    from    the    dipole   torque    formula    (quoted    in
Table-[\ref{t_ORAT}]).  However, an absorption line at $\sim$~1~keV has
been detected  in the  X-ray spectrum of  J1819-1458~\cite{rea09}.  If
this  happens to  be a  cyclotron absorption  line, then  the required
field strengths would be $2 \times 10^{14}$~G and $\sim 10^{11}$~G for
protons and electrons respectively~\cite{mille13}.

It is  also interesting to  consider J1107-5907, which  arguably could
have appeared to be an RRAT had  it been located at a larger distance,
in the  light of  this suggestion.   It is is  an old,  isolated radio
pulsar  and in  the P$_s$-B$_s$  plane located  in the  region between
those of normal and recycled  pulsars. There exists a possibility that
such  pulsars   are  mildly  recycled  in   high-mass  X-ray  binaries
(HMXB)~\cite{konar99a,konar17e}.  If so,  then  RRAT phenomenon  could
indeed be related  to accretion-induced changes in  the magnetic field
structure.

Fig.\ref{f_bp-death}  shows the  possible  import  of the  connections
discussed above.   Not only RRATs  and NPSRs, but  Magnetars along-with
RPSRs   exhibiting   Magnetar    characteristics   have   been   shown
here. Clearly high-magnetic field RRATs inhabit the same region of
the P$_s$-B$_s$ space as the Magnetars.  Now, Magnetars are understood
to be powered  by the decay of their magnetic  fields. Considering the
relative locations, it is quite conceivable that the Magnetars may end
up as  RRATs in the course  of their evolution.

J1107-5907,   J0828-3417,   J1119-6127   and  J1819-1458   have   been
individually marked in Fig.\ref{f_bp-death} to highlight their special
characteristics. It  is seen  that J1107-5907, an  NPSR and  likely to
have RRAT-like behaviour if located  at a larger distance, is somewhat
far away from both the majority of  RRATs as well as NPSRs. Though, we
must remember that only about 25\% of the known RRATs are seen in this
diagram as the rest do not have  a B$_s$ estimate.  On the other hand,
J0828-3417, a similar object, sits right  in the middle of the NPSR as
well as the  RRAT population. Of the two  Magnetar-like objects, while
J1819-1458 is  in the  Magnetar region, J1119-6127  is quite  far away
from all three (RRAT, NPSR, Magnetar) groups.  One possibility is that
these objects  (J1107-5907, J1119-6127) really are  the missing links,
transitioning from one class to another  and are caught at the regions
of transition. Another interesting object is J0835-4510, also shown in
Fig.\ref{f_bp-death},   is  a   known   NPSR   which  sometimes   show
Magnetar-like  behaviour.   These   objects  clearly  demonstrate  the
intertwining connections  between many distinct  observational classes
of neutron stars.

We  have also  shown the  trajectories  of RRATs,  from their  present
location  to  the  death-line,   in  Fig.\ref{f_bp-death}  assuming  a
constant surface  dipolar magnetic  field.  It  is obvious  from these
tracks that,  unless they are  intrinsically different, the  RRATs and
the NPSRs would  be part of the same underlying  population. In \S3 we
have  noted that  they  are  not. Therefore,  it  is  likely that  the
assumption of an RRAT having a constant magnetic  field is incorrect,
even   it   is    true   for   regular   RPSRs    inclusive   of   the
NPSRs~\cite{bhatt92,konar97a}.

Random, sporadic  processes have also  been considered to  explain the
RRAT phenomenon. For example,  the quasi-periodic activity of B1931+24
(J1933+2421) has been  thought to arise from  the interactions between
the  neutron  star  magnetosphere   and  the  precessing  debris  disk
surrounding it~\cite{li06},  or because of the  migration of asteroids
(formed from  supernova fallback material) into  pulsar light-cylinder
region disrupting  the emission  process~\cite{corde08}.  It  has also
been suggested that pulsars can have radiation belts, similar to those
in  planetary magnetosphere,  and sporadic  release of  plasma trapped
therein can  interfere with pulsar  emission processes giving  rise to
RRAT-like behaviour~\cite{luo07}.  Moreover, some  of the single pulse
events  initially  thought  to  be  RRATs  have  now  been  tagged  as
potentially  mislabeled FRBs~\cite{keane16}.  Quite  Clearly, even  if
such explanations hold  good for certain specific  RRATs, the majority
of objects demand a more general scenario.

\section{Conclusions}
The RRATs, one of the latest observational sub-classes, comprise a tiny
($< 5\%$) subset of the known neutron stars. As yet, there is no clear
understanding of the nature of their sporadic emission.  In this work,
we have considered  the population of RRATs and discussed  some of the
prominent hypotheses (offered  to explain the RRAT  phenomenon) in the
light of the general properties of this population.

First of all,  considering the population characteristics,  we come to
the conclusion  that the  RRAT behaviour can  neither be  explained as
special  observational  effects  or  as a  result  of  certain  random
processes.

One of the main trains of  thought, regarding the RRATs has been about
their connection  with the NPSRs.   Comparing the two  populations, we
find that -
\bei
   \i The RRATS are bounded by death-line~{\bf[2]}, like the NPSRs.
   \i RRATs have, on the average, higher P$_s$ and B$_s$ than NPSRs.
   \i RRATs tend to have smaller DM compared to the NPSRs.
   \i  P$_s$ \&  B$_s$ of  RRATs  and NPSRs  as well  as the  combined
   population appear to have log-normal distribution.
   \i The statistical  distributions of P$_s$, B$_s$ and  DM for RRATs
   and  NPSRs  do  not  appear   to  come  from  the  same  underlying
   distribution.
\eei
Thus,  we conclude  that the  RRATs and  the NPSRs, even if  connected
through  evolution, do  not  come directly  from  the same  underlying
population, and  RRAT behaviour is  unlikely to be a  manifestation of
extreme nulling.

Another important conjecture has been  to attribute the RRAT behaviour
to the  proximity of these neutron  stars to the death-line.   We have
considered  this  assertion  by quantifying  the  death-line  nearness
through the  proximity parameter, $\epsilon$.  We find that  the RRATs
appear to have a significant part of their active life left before they
would  reach  the  death-line.   Even though  this  finding  crucially
depends on the  assumption of a constant magnetic field,  if true this
indicates that the  RRAT behaviour is not a direct  consequence of old
age.

However, we find two very different and intriguing conjectures to hold
some promise.   First, the  RRAT phenomenon  could be  associated with
re-structuring   of   magnetic   fields.    This   could   either   be
accretion-induced, through the revival of  `dead' pulsars. Or it could
be  glitch-induced.  Second,  the  RRATs could  be later  evolutionary
stages   of    other   varieties    of   neutron   stars    like   the
Magnetars~\cite{konar12}.  Even though a recent work suggests that the
Magnetars  are  likely  to  evolve into  XDINS~\cite{jawor22},  in  an
independent work~\cite{chowh22}  we find that even  though majority of
the Magnetars are  likely to evolve into XDINS, some  of the Magnetars
as  well as  most of  the high  magnetic radio  pulsars are  likely to
evolve into RRATs.

\section{Acknowledgment}
We thank Priya S Hasan \& S N Hasan whose 2020 workshop, on ``Astronomy
from Archival  Data" as part of  an IAU-OAD project, helped  bring the
authors of this work together.

\label{lastpage}

\bibliography{mnrasmnemonic,adsrefs}

\begin{thebibliography}{}

\bibitem[\protect\citeauthoryear{{Archibald} et~al.}{{Archibald}
  et~al.}{2017}]{archi17}
{Archibald} R.~F. et~al., 2017, \apjl, 849, L20

\bibitem[\protect\citeauthoryear{{Backer}}{{Backer}}{1970}]{backe70}
{Backer} D.~C., 1970, \nat, 228, 42

\bibitem[\protect\citeauthoryear{{Bhattacharya} et~al.}{{Bhattacharya}
  et~al.}{1992}]{bhatt92}
{Bhattacharya} D., {Wijers} R.~A.~M.~J., {Hartman} J.~W.,  {Verbunt} F., 1992,
  \aap, 254, 198

\bibitem[\protect\citeauthoryear{{Bhattacharyya} et~al.}{{Bhattacharyya}
  et~al.}{2018}]{bhatt18}
{Bhattacharyya} B. et~al., 2018, \mnras, 477, 4090

\bibitem[\protect\citeauthoryear{{Burke-Spolaor} \& {Bailes}}{{Burke-Spolaor}
  \& {Bailes}}{2010}]{burke10}
{Burke-Spolaor} S.,  {Bailes} M., 2010, \mnras, 402, 855

\bibitem[\protect\citeauthoryear{{Burke-Spolaor} et~al.}{{Burke-Spolaor}
  et~al.}{2011}]{burke11}
{Burke-Spolaor} S. et~al., 2011, \mnras, 416, 2465

\bibitem[\protect\citeauthoryear{{Burke-Spolaor} et~al.}{{Burke-Spolaor}
  et~al.}{2012}]{burke12}
{Burke-Spolaor} S. et~al., 2012, \mnras, 423, 1351

\bibitem[\protect\citeauthoryear{{Chen} \& {Ruderman}}{{Chen} \&
  {Ruderman}}{1993}]{chen93a}
{Chen} K.,  {Ruderman} M., 1993, \apj, 402, 264

\bibitem[\protect\citeauthoryear{{Chowhan}, {Konar}, \& {Banik}}{{Chowhan}
  et~al.}{2022}]{chowh22}
{Chowhan} T.~T., {Konar} S.,  {Banik} S., 2022, in {Troja} E.,  {Baring} M.,
  ed, Neutron Star Astrophysics at the Crossroads: Magnetars and the
  Multimessenger Revolution, Proceedings of IAU Symposium No. 363 ({\bf \em in
  press}), arXiv:2201.05248

\bibitem[\protect\citeauthoryear{{Cordes} \& {Shannon}}{{Cordes} \&
  {Shannon}}{2008}]{corde08}
{Cordes} J.~M.,  {Shannon} R.~M., 2008, \apj, 682, 1152

\bibitem[\protect\citeauthoryear{{Cui} et~al.}{{Cui} et~al.}{2017}]{cui17}
{Cui} B.~Y., {Boyles} J., {McLaughlin} M.~A.,  {Palliyaguru} N., 2017, \apj,
  840, 5

\bibitem[\protect\citeauthoryear{{Deneva} et~al.}{{Deneva}
  et~al.}{2013}]{denev13}
{Deneva} J.~S., {Stovall} K., {McLaughlin} M.~A., {Bates} S.~D., {Freire}
  P.~C.~C., {Martinez} J.~G., {Jenet} F.,  {Bagchi} M., 2013, \apj, 775, 51

\bibitem[\protect\citeauthoryear{{Esamdin} et~al.}{{Esamdin}
  et~al.}{2012}]{esamd12}
{Esamdin} A., {Abdurixit} D., {Manchester} R.~N.,  {Niu} H.~B., 2012, \apjl,
  759, L3

\bibitem[\protect\citeauthoryear{{Gen{\c{c}}ali} \& {Ertan}}{{Gen{\c{c}}ali} \&
  {Ertan}}{2018}]{genca18}
{Gen{\c{c}}ali} A.~A.,  {Ertan} {\"U}., 2018, \mnras, 481, 244

\bibitem[\protect\citeauthoryear{{Good} et~al.}{{Good} et~al.}{2021}]{good21}
{Good} D.~C. et~al., 2021, \apj, 922, 43

\bibitem[\protect\citeauthoryear{{Hessels} et~al.}{{Hessels}
  et~al.}{2008}]{hesse08}
{Hessels} J.~W.~T., {Ransom} S.~M., {Kaspi} V.~M., {Roberts} M.~S.~E.,
  {Champion} D.~J.,  {Stappers} B.~W., 2008, in American Institute of Physics
  Conference Series, Vol. 983, {Bassa} C., {Wang} Z., {Cumming} A.,  {Kaspi}
  V.~M., ed, 40 Years of Pulsars: Millisecond Pulsars, Magnetars and More, p.
  613

\bibitem[\protect\citeauthoryear{{Hewish} et~al.}{{Hewish}
  et~al.}{1968}]{hewis68}
{Hewish} A., {Bell} S.~J., {Pilkington} J.~D.~H., {Scott} P.~F.,  {Collins}
  R.~A., 1968, \nat, 217, 709

\bibitem[\protect\citeauthoryear{{Jawor} \& {Tauris}}{{Jawor} \&
  {Tauris}}{2022}]{jawor22}
{Jawor} J.~A.,  {Tauris} T.~M., 2022, \mnras, 509, 634

\bibitem[\protect\citeauthoryear{{Kaplan} et~al.}{{Kaplan}
  et~al.}{2009}]{kapla09b}
{Kaplan} D.~L., {Esposito} P., {Chatterjee} S., {Possenti} A., {McLaughlin}
  M.~A., {Camilo} F., {Chakrabarty} D.,  {Slane} P.~O., 2009, \mnras, 400, 1445

\bibitem[\protect\citeauthoryear{{Kaspi}}{{Kaspi}}{2010}]{kaspi10}
{Kaspi} V.~M., 2010, Proceedings of the National Academy of Science, 107, 7147

\bibitem[\protect\citeauthoryear{{Keane}}{{Keane}}{2016}]{keane16}
{Keane} E.~F., 2016, \mnras, 459, 1360

\bibitem[\protect\citeauthoryear{{Keane} \& {Kramer}}{{Keane} \&
  {Kramer}}{2008}]{keane08}
{Keane} E.~F.,  {Kramer} M., 2008, \mnras, 391, 2009

\bibitem[\protect\citeauthoryear{{Keane} et~al.}{{Keane}
  et~al.}{2011}]{keane11a}
{Keane} E.~F., {Kramer} M., {Lyne} A.~G., {Stappers} B.~W.,  {McLaughlin}
  M.~A., 2011, \mnras, 415, 3065

\bibitem[\protect\citeauthoryear{{Keane} et~al.}{{Keane}
  et~al.}{2010}]{keane10b}
{Keane} E.~F., {Ludovici} D.~A., {Eatough} R.~P., {Kramer} M., {Lyne} A.~G.,
  {McLaughlin} M.~A.,  {Stappers} B.~W., 2010, \mnras, 401, 1057

\bibitem[\protect\citeauthoryear{{Keane} \& {McLaughlin}}{{Keane} \&
  {McLaughlin}}{2011}]{keane11b}
{Keane} E.~F.,  {McLaughlin} M.~A., 2011, Bulletin of the Astronomical Society
  of India, 39, 333

\bibitem[\protect\citeauthoryear{{Knight} et~al.}{{Knight}
  et~al.}{2006}]{knigh06}
{Knight} H.~S., {Bailes} M., {Manchester} R.~N., {Ord} S.~M.,  {Jacoby} B.~A.,
  2006, \apj, 640, 941

\bibitem[\protect\citeauthoryear{{Konar}}{{Konar}}{2012}]{konar12}
{Konar} S., 2012, in COSPAR Meeting, Vol.~39, 39th COSPAR Scientific Assembly,
  p. 961

\bibitem[\protect\citeauthoryear{{Konar}}{{Konar}}{2013}]{konar13}
{Konar} S., 2013, in Astronomical Society of India Conference Series, Vol.~8,
  {Das} S., {Nandi} A.,  {Chattopadhyay} I., ed, Astronomical Society of India
  Conference Series, p.~89

\bibitem[\protect\citeauthoryear{{Konar}}{{Konar}}{2017}]{konar17e}
{Konar} S., 2017, Journal of Astrophysics and Astronomy, 38, 47

\bibitem[\protect\citeauthoryear{{Konar} et~al.}{{Konar}
  et~al.}{2016}]{konar16c}
{Konar} S. et~al., 2016, Journal of Astrophysics and Astronomy, 37, 36

\bibitem[\protect\citeauthoryear{{Konar} \& {Bhattacharya}}{{Konar} \&
  {Bhattacharya}}{1997}]{konar97a}
{Konar} S.,  {Bhattacharya} D., 1997, \mnras, 284, 311

\bibitem[\protect\citeauthoryear{{Konar} \& {Bhattacharya}}{{Konar} \&
  {Bhattacharya}}{1999}]{konar99a}
{Konar} S.,  {Bhattacharya} D., 1999, \mnras, 303, 588

\bibitem[\protect\citeauthoryear{{Konar} \& {Deka}}{{Konar} \&
  {Deka}}{2019}]{konar19e}
{Konar} S.,  {Deka} U., 2019, Journal of Astrophysics and Astronomy, 40, 42

\bibitem[\protect\citeauthoryear{{Li}}{{Li}}{2006}]{li06}
{Li} X.-D., 2006, \apjl, 646, L139

\bibitem[\protect\citeauthoryear{{Logvinenko}, {Tyul'bashev}, \&
  {Malofeev}}{{Logvinenko} et~al.}{2020}]{logvi20}
{Logvinenko} S.~V., {Tyul'bashev} S.~A.,  {Malofeev} V.~M., 2020, Bulletin of
  the Lebedev Physics Institute, 47, 390

\bibitem[\protect\citeauthoryear{{Lu} et~al.}{{Lu} et~al.}{2019}]{lu19}
{Lu} J. et~al., 2019, Science China Physics, Mechanics, and Astronomy, 62,
  959503

\bibitem[\protect\citeauthoryear{{Luo} \& {Melrose}}{{Luo} \&
  {Melrose}}{2007}]{luo07}
{Luo} Q.,  {Melrose} D., 2007, \mnras, 378, 1481

\bibitem[\protect\citeauthoryear{{Lyne} et~al.}{{Lyne} et~al.}{2009}]{lyne09}
{Lyne} A.~G., {McLaughlin} M.~A., {Keane} E.~F., {Kramer} M., {Espinoza} C.~M.,
  {Stappers} B.~W., {Palliyaguru} N.~T.,  {Miller} J., 2009, \mnras, 400, 1439

\bibitem[\protect\citeauthoryear{{Manchester} et~al.}{{Manchester}
  et~al.}{2005}]{manch05b}
{Manchester} R.~N., {Hobbs} G.~B., {Teoh} A.,  {Hobbs} M., 2005, VizieR Online
  Data Catalog, 7245, 0

\bibitem[\protect\citeauthoryear{{McLaughlin} et~al.}{{McLaughlin}
  et~al.}{2009}]{mclau09}
{McLaughlin} M.~A. et~al., 2009, \mnras, 400, 1431

\bibitem[\protect\citeauthoryear{{McLaughlin} et~al.}{{McLaughlin}
  et~al.}{2006}]{mclau06}
{McLaughlin} M.~A. et~al., 2006, \nat, 439, 817

\bibitem[\protect\citeauthoryear{{McLaughlin} et~al.}{{McLaughlin}
  et~al.}{2007}]{mclau07}
{McLaughlin} M.~A. et~al., 2007, \apj, 670, 1307

\bibitem[\protect\citeauthoryear{{Miller} et~al.}{{Miller}
  et~al.}{2013}]{mille13}
{Miller} J.~J., {McLaughlin} M.~A., {Rea} N., {Lazaridis} K., {Keane} E.~F.,
  {Kramer} M.,  {Lyne} A., 2013, \apj, 776, 104

\bibitem[\protect\citeauthoryear{{Popov}, {Turolla}, \& {Possenti}}{{Popov}
  et~al.}{2006}]{popov06}
{Popov} S.~B., {Turolla} R.,  {Possenti} A., 2006, \mnras, 369, L23

\bibitem[\protect\citeauthoryear{{Rea} et~al.}{{Rea} et~al.}{2010}]{rea10}
{Rea} N. et~al., 2010, \mnras, 407, 1887

\bibitem[\protect\citeauthoryear{{Rea} et~al.}{{Rea} et~al.}{2009}]{rea09}
{Rea} N. et~al., 2009, \apjl, 703, L41

\bibitem[\protect\citeauthoryear{{Ritchings}}{{Ritchings}}{1976}]{ritch76}
{Ritchings} R.~T., 1976, \mnras, 176, 249

\bibitem[\protect\citeauthoryear{{Ruderman} \& {Sutherland}}{{Ruderman} \&
  {Sutherland}}{1975}]{ruder75}
{Ruderman} M.~A.,  {Sutherland} P.~G., 1975, \apj, 196, 51

\bibitem[\protect\citeauthoryear{{Shitov} et~al.}{{Shitov}
  et~al.}{2009}]{shito09}
{Shitov} Y.~P., {Kuzmin} A.~D., {Dumskii} D.~V.,  {Losovsky} B.~Y., 2009,
  Astronomy Reports, 53, 561

\bibitem[\protect\citeauthoryear{{Tyul'bashev} et~al.}{{Tyul'bashev}
  et~al.}{2021}]{tyulb21}
{Tyul'bashev} S., {Kitaeva} M., {Logvinenko} S.,  G.E. T., 2021, Astronomy
  Reports (in press)

\bibitem[\protect\citeauthoryear{{Tyul'bashev}, {Tyul'bashev}, \&
  {Malofeev}}{{Tyul'bashev} et~al.}{2018}]{tyulb18}
{Tyul'bashev} S.~A., {Tyul'bashev} V.~S.,  {Malofeev} V.~M., 2018, \aap, 618,
  A70

\bibitem[\protect\citeauthoryear{{von Mises}}{{von Mises}}{1980}]{mises64}
{von Mises} R., 1980, {Mathematical Theory of Probability and Statistics}.
\newblock Academic Press, New York

\bibitem[\protect\citeauthoryear{{Weltevrede}, {Johnston}, \&
  {Espinoza}}{{Weltevrede} et~al.}{2011}]{welte11}
{Weltevrede} P., {Johnston} S.,  {Espinoza} C.~M., 2011, \mnras, 411, 1917

\bibitem[\protect\citeauthoryear{{Weltevrede} et~al.}{{Weltevrede}
  et~al.}{2006}]{welte06}
{Weltevrede} P., {Stappers} B.~W., {Rankin} J.~M.,  {Wright} G.~A.~E., 2006,
  \apjl, 645, L149

\bibitem[\protect\citeauthoryear{{Young} et~al.}{{Young}
  et~al.}{2014}]{young14}
{Young} N.~J., {Weltevrede} P., {Stappers} B.~W., {Lyne} A.~G.,  {Kramer} M.,
  2014, \mnras, 442, 2519

\end{thebibliography}
\bibliographystyle{mnras}

\appendix

\onecolumn

\section{The RRATs}
\label{append}
\begin{minipage}{0.95\textwidth}
The following tables list the known RRATs, detected till date, that we have used for our calculations in this work. Majority of the sources are obtained from the {\bf RRATalog} site maintained by Bingyi Cui and Maura McLaughlin. For the rest, discovery papers have been referred to. The parameter values for spin period (P$_s$), dispersion measure (DM), characteristic age ($\tau_c$) and surface dipolar field (B$_s$) are taken from the {\bf ATNF Pulsar Catalogue}~\cite{manch05b}, except where the objects are not yet included in the ATNF list. Parameter values for these second set of objects are taken from the discovery papers and the references are marked with a `{\bf -P}'. For some sources, different names have been used by different groups. We have primarily used the ATNF names and indicated the alternative names in the {\em `Other-Name'} column. The references cited in the tables with numbers ranging from 1 to 8 correspond to the following. \\

{\bf (1)} - RRATalog (last update September, 2016); \\
{\bf (2)} - \citeN{welte11}; \\
{\bf (3)} - \citeN{esamd12}; \\
{\bf (4)} - \citeN{denev13}; \\
{\bf (5)} - \citeN{cui17}; \\
{\bf (6)} - \citeN{tyulb18}; \\ 
{\bf (7)} - \citeN{good21}; \\ 
{\bf (8)} - \citeN{logvi20}; \\ 
{\bf (9)} - \citeN{tyulb21}. \\
 
{\bf RRATalog} : \underline{\tt http://astro.phys.wvu.edu/rratalog/}  \\

{\bf ATNF} : \underline{\tt http://www.atnf.csiro.au/research/pulsar/psrcat/} 
\end{minipage}


%
\begin{table*} 
\caption{List of RRATs}
\label{t_list01}
\begin{minipage}{0.95\textwidth}
\centering
\begin{tabular}{|l|l|l|l|l|l|l|l|l|} \toprule
     &  B-Name     & J-Name      & Other-Name  & P$_s$ &  DM    &  $\tau$    & B$_s$    &        \\
     &             &             &             &  s    & Pfc.cm$^{-3}$ & yr    & G        &        \\ \midrule
 001 & J0054+66    & J0054+66    &             &  1.39 &  15.00 &            &          &  {\bf (1)}   \\
 002 & J0054+69    & J0054+69    &             &       &  90.30 &            &          &  {\bf (1)}   \\
 003 & J0103+54    & J0103+54    &             &  0.35 &  55.60 &            &          &  {\bf (1)}   \\
 004 & J0121+53    & J0121+53    &             &  2.72 &  91.38 &            &          &  {\bf (7)}   \\
 005 & J0139+3336  & J0139+3336  &             &  1.25 &  21.23 &  9.58e+06  & 1.62e+12 &  {\bf (6)}   \\
 006 & J0156+04    & J0156+04    &             &       &  27.50 &            &          &  {\bf (1)}   \\
 007 & J0318+1341  & J0318+1341  &             &  1.97 &  12.05 &            &          &  {\bf (1-P)} \\
 008 & J0201+7005  & J0201+7005  &             &  1.35 &  21.03 &  3.88e+06  & 2.76e+12 &  {\bf (1)}   \\
 009 & J0302+2252  & J0302+2252  &  J0301+20   &  1.21 &  18.99 &  2.32e+08  & 3.19e+11 &  {\bf (1)}   \\
 010 & J0305+4001  & J0305+4001  &             &       &  24.00 &            &          &  {\bf (6)}   \\
 011 & J0332+79    & J0332+79    &             &  2.06 &  16.67 &            &          &  {\bf (1)}   \\
 012 & J0410-31    & J0410-31    &             &  1.88 &   9.20 &            &          &  {\bf (1)}   \\
 013 & J0441-04    & J0441-04    &             &       &  20.00 &            &          &  {\bf (1)}   \\
 014 & J0447-04    & J0447-04    &             &  2.19 &  29.83 &            &          &  {\bf (1)}   \\ 
 015 & J0452+1651  & J0452+1651  &             &       &  19.00 &            &          &  {\bf (6)}   \\
 016 & J0513-04    & J0513-04    &             &       &  18.50 &            &          &  {\bf (1)}   \\
 017 & J0534+34    & J0534+34    & J0534+3407  &       &  24.50 &            &          &  {\bf (6)}   \\
 018 & J0544+20    & J0544+20    & J0544-20    &       &  56.90 &            &          &  {\bf (1)}   \\
 019 & J0545-03    & J0545-03    &             &  1.07 &  67.20 &            &          &  {\bf (1)}   \\
 020 & J0550+09    & J0550+09    &             &  1.75 &  86.60 &            &          &  {\bf (1)}   \\
 021 & J0609+1635  & J0609+1635  &             &       &  85.00 &            &          &  {\bf (6)}   \\
 022 & J0614-03    & J0614-03	 &             &  0.14 &  17.90 &            &          &  {\bf (1-P)} \\ 
 023 & J0621-55    & J0621-55    &             &       &  22.00 &            &          &  {\bf (1)}   \\
 024 & J0625+1730  & J0625+1730  &             &       &  58.00 &            &          &  {\bf (6)}   \\
 025 & J0627+16    & J0627+16    &             &  2.18 & 113.00 &            &          &  {\bf (1)}   \\
 026 & J0628+0909  & J0628+0909  &  J0628+09   &  1.24 &  88.30 &  3.59e+07  & 8.35e+11 &  {\bf (1)}   \\
 027 & J0640+0744  & J0640+0744  &  J0641+07   &       &  52.00 &            &          &  {\bf (6)}   \\ 
 028 & J0736-6304  & J0736-6304  &  J0735-62   &  4.86 &  19.40 &  5.07e+05  & 2.75e+13 &  {\bf (1,5-P)} \\
 029 & J0803+34    & J0803+34    &  J0803+3410 &       &  34.00 &            &          &  {\bf (5)}   \\
 030 & J0812+8626  & J0812+8626  &             &       &  40.25 &            &          &  {\bf (9)}   \\
 031 & B0826-34    & J0828-3417  &             & 1.85  & 52.20  &  2.94e+07  & 1.37e+12 &  {\bf (3)}   \\
 032 & J0837-24    & J0837-24    &             &       & 142.80 &            &          &  {\bf (1)}   \\
 033 & J0845-36    & J0845-36    &             &  0.21 &  29.00 &  2.61e+07  & 1.67e+11 &  {\bf (1-P)} \\
 034 & J0847-4316  & J0847-4316  &             &  5.98 & 292.50 &  7.90e+05  & 2.71e+13 &  {\bf (1)}   \\
 035 & J0912-3851  & J0912-3851  &  J0912-38   &  1.53 &  71.50 &  6.74e+06  & 2.37e+12 &  {\bf (1)}   \\
 036 & J0923-31    & J0923-31    &             &       &  72.00 &            &          &  {\bf (1)}   \\ 
 037 & J0941+1621  & J0941+1621  &             &       &  23.00 &            &          &  {\bf (6)}   \\
 038 & J0941-39    & J0941-39    &             &  0.59 &  78.20 &            &          &  {\bf (1)}   \\
 039 & J0957-06    & J0957-06    &             &  1.72 &  26.95 &            &          &  {\bf (1)}   \\
 040 & J1005+30    & J1005+30    &             &       &  17.50 &            &          &  {\bf (6)}   \\ 
 041 & J1010+15    & J1010+15    &             &       &  42.00 &            &          &  {\bf (4)}   \\
 042 & J1014-48    & J1014-48    &             &  1.51 &  87.00 &            &          &  {\bf (1)}   \\
 043 & J1048-5838  & J1048-5838  &             &  1.23 &  70.70 &  1.60e+06  & 3.92e+12 &  {\bf (1)}   \\
 044 & J1059-01    & J1059-01    &             &       &  18.70 &            &          &  {\bf (1)}   \\
 045 & J1111-55    & J1111-55    &             &       & 235.00 &            &          &  {\bf (1)}   \\
 046 & J1119-6127  & J1119-6127  &             &  0.41 & 704.80 & 1.61e+03   & 4.10e+13 &  {\bf (2)}   \\
 047 & J1126-27    & J1126-27    &             &  0.36 &  26.86 &            &          &  {\bf (1)}   \\
 048 & J1129-53    & J1129-53    &             &  1.06 &  77.00 &            &          &  {\bf (1)}   \\
 049 & J1132+0921  & J1132+0921  &             &       &  22.00 &            &          &  {\bf (6)}   \\
 050 & J1132+25    & J1132+25    &  J1132+2515 &  1.00 &  23.00 &            &          &  {\bf (6)}   \\ \bottomrule
\end{tabular}
\end{minipage}
\label{t_rrat1}
\end{table*}
\begin{table*} 
\caption{List of RRATs - continued.}
\label{t_list02}
\begin{minipage}{0.95\textwidth}
\centering
\begin{tabular}{|l|l|l|l|l|l|l|l|l|} \toprule
     &  B-Name     & J-Name      & Other-Name  & P$_s$ &  DM    &  $\tau$    & B$_s$    &        \\
     &             &             &             &  s    & PVC.cm$^{-3}$ & yr    & G        &        \\ \midrule
 051 & J1135-49    & J1135-49    &             &       & 114.00 &            &          &  {\bf (1)}   \\ 
 052 & J1153-21    & J1153-21    &             &  2.34 &  34.80 &            &          &  {\bf (1)}   \\
 053 & J1216-50    & J1216-50    &             &  6.35 & 110.00 &            &          &  {\bf (1)}   \\
 054 & J1226-3223  & J1226-3223  &             &  6.19 &  36.70 &  1.39e+07  & 6.69e+12 &  {\bf (1)}   \\
 055 & J1252+53    & J1252+53    &             &  0.22 &  20.70 &            &          &  {\bf (7)}   \\
 056 & J1307-67    & J1307-67    &  J1308-67   &  3.65 &  44.00 &            &          &  {\bf (1)}   \\
 057 & J1311-59    & J1311-59    &             &       & 152.00 &            &          &  {\bf (1)}   \\
 058 & J1317-5759  & J1317-5759  &             &  2.64 & 145.30 &  3.33e+06  & 5.83e+12 &  {\bf (1)}   \\
 059 & J1326+33    & J1326+33    &             &  0.04 &   4.00 &            &          &  {\bf (6)}   \\
 060 & J1329+13    & J1329+13    & J1329+1349  &       &  12.00 &            &          &  {\bf (6)}   \\
 061 & J1332-03    & J1332-03    &             &  1.11 &  27.10 &            &          &  {\bf (1)}   \\
 062 & J1336-20    & J1336-20    &             &  0.18 &  19.30 &            &          &  {\bf (1-P)} \\
 063 & J1336+33    & J1336+33    &  J1336+3346 &  3.01 &   8.50 &            &          &  {\bf (6)}   \\
 064 & J1346+0622  & J1346+0622  &             &       &   8.00 &            &          &  {\bf (6)}   \\
 065 & J1354+24    & J1354+24    &             &       &  20.00 &            &          &  {\bf (1)}   \\
 066 & J1400+21    & J1400+21    & J1400+2127  &       &  10.50 &            &          &  {\bf (6)}   \\
 067 & J1404+1210  & J1404+1210  &             &  2.65 &  17.05 &            &          &  {\bf (6-P)} \\
 068 & J1404-58    & J1404-58    &             &       & 229.00 &            &          &  {\bf (1)}   \\
 069 & J1424-56    & J1424-56    &  J1423-56   &  1.43 &  32.90 &            &          &  {\bf (1)}   \\
 070 & J1433+00    & J1433+00    &             &       &  23.50 &            &          &  {\bf (1)}   \\
 071 & J1439+76    & J1439+76    &             &  0.95 &  22.29 &            &          &  {\bf (1)}   \\
 072 & J1444-6026  & J1444-6026  &             &  4.76 & 367.70 &  4.07e+06  & 9.51e+12 &  {\bf (1)}   \\
 073 & J1502+28    & J1502+28    &  J1502+2813 &  3.78 &  14.00 &            &          &  {\bf (6)}   \\
 074 & J1513-5946  & J1513-5946  &             &  1.05 & 171.70 &  1.94e+06  & 3.02e+12 &  {\bf (1)}   \\
 075 & J1534-46    & J1534-46    &             &  0.36 &  64.40 &            &          &  {\bf (1)}   \\
 076 & J1538+2345  & J1538+2345  &             &  3.45 &  14.91 &  7.93e+06  & 4.93e+12 &  {\bf (1)}   \\
 077 & J1541-42    & J1541-42    &             &       &  60.00 &            &          &  {\bf (1)}   \\
 078 & J1549-57    & J1549-57    &             &  0.74 &  17.70 &            &          &  {\bf (1)}   \\
 079 & J1550+0943  & J1550+0943  &             &       &  21.00 &            &          &  {\bf (8)}   \\
 080 & J1554+18    & J1554+18    &             &       &  24.00 &            &          &  {\bf (1)}   \\
 081 & J1554-5209  & J1554-5209  &             &  0.13 & 130.80 &  8.65e+05  & 5.42e+11 &  {\bf (1)}   \\
 082 & J1555+01    & J1555+01    & J1555+0108  &       &  18.50 &            &          &  {\bf (6)}   \\
 083 & J1603+18    & J1603+18    &             &  0.50 &  29.70 &            &          &  {\bf (1)}   \\
 084 & J1610-17    & J1610-17    &             &  1.30 &  52.50 &            &          &  {\bf (1-P)} \\
 085 & J1611-01    & J1611-01    &             &  1.30 &  27.21 &            &          &  {\bf (1)}   \\
 086 & J1623-0841  & J1623-0841  &  J1623-08   &  0.50 &  59.79 &  4.08e+06  & 1.00e+12 &  {\bf (1)}   \\ 
 087 & J1647-3607  & J1647-3607  &  J1647-36   &  0.21 & 224.00 &  2.61e+07  & 1.67e+11 &  {\bf (1)}   \\ 
 088 & J1649-4653  & J1649-4653  &  J1649-46   &  0.56 & 331.00 &  1.78e+05  & 5.32e+12 &  {\bf (1)}   \\
 089 & J1652-4406  & J1652-4406  &             &  7.71 & 786.00 &  1.29e+07  & 8.66e+12 &  {\bf (1)}   \\
 090 & J1654-23    & J1654-23    &  J1653-2330 &  0.55 &  74.50 &  6.31e+05  & 1.62e+12 &  {\bf (1-P)} \\
 091 & J1703-38    & J1703-38    &             &  6.44 & 375.00 &            &          &  {\bf (1)}   \\
 092 & J1705-04    & J1705-04    &             &  0.24 &  42.95 &            &          &  {\bf (1)}   \\
 093 & J1707-4417  & J1707-4417  &  J1704-44   &  5.76 & 380.00 &  7.84e+06  & 8.29e+12 &  {\bf (1)}   \\
 094 & J1709-43    & J1709-43    &             &       & 228.00 &            &          &  {\bf (1)}   \\
 095 & J1717+03    & J1717+03    &             &  3.90 &  25.60 &            &          &  {\bf (1)}   \\
 096 & J1720+00    & J1720+00    &             &  3.36 &  46.20 &            &          &  {\bf (1)}   \\
 097 & J1724-35    & J1724-35    &             &  1.42 & 554.90 &            &          &  {\bf (1)}   \\
 098 & J1727-29    & J1727-29    &             &       &  93.00 &            &          &  {\bf (1)}   \\
 099 & J1732+2700  & J1732+2700  &             &       &  36.50 &            &          &  {\bf (6)}   \\ 
 100 & J1739-2521  & J1739-2521  &  J1739-25   &  1.82 & 186.40 &  1.20e+08  & 6.69e+11 &  {\bf (1)}   \\ \bottomrule
\end{tabular}
\end{minipage}
\label{t_rrat2}
\end{table*}
\begin{table*} 
\caption{List of RRATs - continued.}
\label{t_list03}
\begin{minipage}{0.95\textwidth}
\centering
\begin{tabular}{|l|l|l|l|l|l|l|l|l|} \toprule
     &  B-Name     & J-Name      & Other-Name  & P$_s$ &  DM    &  $\tau$    & B$_s$    &        \\
     &             &             &             &  s    & PRC.cm$^{-3}$ & yr    & G        &        \\ \midrule
 101 & J1753-12    & J1753-12    &             &  0.40 &  73.20 &            &          &  {\bf (1)}   \\ 
 102 & J1753-38    & J1753-38    &             &  0.67 & 168.00 &            &          &  {\bf (1)}   \\
 103 & J1754-3014  & J1754-3014  &  J1754-30   &  1.32 &  89.70 &  4.72e+06  & 2.45e+12 &  {\bf (1)}   \\
 104 & J1807-2557  & J1807-2557  &  J1807-25   &  2.76 & 385.00 &  8.77e+06  & 3.76e+12 &  {\bf (1)}   \\
 105 & J1819-1458  & J1819-1458  &             &  4.26 & 196.00 &  1.20e+05  & 4.96e+13 &  {\bf (1)}   \\
 106 & J1825-33    & J1825-33    &             &  1.27 &  43.20 &            &          &  {\bf (1)}   \\
 107 & J1826-1419  & J1826-1419  &             &  0.77 & 160.00 &  1.39e+06  & 2.63e+12 &  {\bf (1)}   \\
 108 & J1838+50    & J1838+50    &             &  2.58 &  21.81 &            &          &  {\bf (7)}   \\
 109 & J1839-0141  & J1839-0141  &  J1839-01   &  0.93 & 293.20 &  2.49e+06  & 2.38e+12 &  {\bf (1)}   \\
 110 & J1840-1419  & J1840-1419  &             &  6.59 &  19.40 &  1.65e+07  & 6.55e+12 &  {\bf (1)}   \\
 111 & J1841-04    & J1841-04    & J1841-0448  &       &  29.00 &            &          &  {\bf (6)}   \\
 112 & J1843+01    & J1843+01    &             &  1.27 & 251.90 &            &          &  {\bf (1)}   \\
 113 & J1846-0257  & J1846-0257  &             &  4.48 & 237.00 &  4.42e+05  & 2.71e+13 &  {\bf (1)}   \\
 114 & J1848-1243  & J1848-1243  &  J1842-12   &  0.42 &  91.96 &  1.49e+07  & 4.32e+11 &  {\bf (1)}   \\
 115 & J1848+1516  & J1848+1516  &  J1849+15   &  2.24 &  77.44 &  2.11e+07  & 1.96e+12 &  {\bf (1)}   \\
     &             &             &  J1848+1518 &       &  75.00 &            &          &  {\bf (6)}   \\
 116 & J1849+0112  & J1849+0112  &             &  1.83 & 217.20 &  1.59e+06  & 5.01e+12 &  {\bf (1-P)} \\
 117 & J1850+15    & J1850+15    &             &  1.38 &  24.70 &            &          &  {\bf (1)}   \\
 118 & J1853+04    & J1853+04    &             &  1.32 & 549.30 &            &          &  {\bf (1)}   \\
 119 & J1854+0306  & J1854+0306  &             &  4.56 & 192.40 &  4.98e+05  & 2.60e+13 &  {\bf (1)}   \\
 120 & J1854-1557  & J1854-1557  &             &  3.45 & 150.00 &  1.21e+07  & 4.00e+12 &  {\bf (1)}   \\
 121 & J1856+09    & J1856+09    &             &  2.17 & 193.40 &            &          &  {\bf (1)}   \\
 122 & J1859+07    & J1859+07    &             &       & 303.00 &            &          &  {\bf (1)}   \\
 123 & J1901+11    & J1901+11    &             &  0.41 & 268.90 &            &          &  {\bf (1)}   \\
 124 & J1905+0414  & J1905+0414  &             &       & 383.00 &            &          &  {\bf (1)}   \\
 125 & J1905+0902  & J1905+0902  &  J1905+09   &  0.22 & 433.40 &  9.88e+05  & 8.84e+11 &  {\bf (1)}   \\
 126 & J1906+03    & J1906+03    &             &  1.26 & 212.00 &            &          &  {\bf (1-P)} \\
 127 & J1909+0641  & J1909+0641  &  J1909+06   &  0.74 &  36.70 &  3.65e+06  & 1.56e+12 &  {\bf (1)}   \\
 128 & J1911+00    & J1911+00    &             &  6.94 & 100.00 &            &          &  {\bf (1)}   \\
 129 & J1912+08    & J1912+08    &             &       &  96.00 &            &          &  {\bf (1)}   \\
 130 & J1913+1330  & J1913+1330  &             &  0.92 & 175.64 &  1.69e+06  & 2.86e+12 &  {\bf (1)}   \\
 131 & J1915+06    & J1915+06    &             &  0.64 & 214.50 &            &          &  {\bf (1)}   \\
 132 & J1915-11    & J1915-11    &             &  2.18 &  91.06 &            &          &  {\bf (1)}   \\
 133 & J1917+11    & J1917+11	 &             &  5.06 & 319.00 &            &          &  {\bf (6-P)} \\
 134 & J1917+1723  & J1917+1723  &             &       &  38.00 &            &          &  {\bf (6)}   \\
 135 & J1919+1745  & J1919+1745  &  J1919+17   &  2.08 & 142.30 &  1.93e+07  & 1.91e+12 &  {\bf (1)}   \\
 136 & J1925-16    & J1925-16    &             &  3.89 &  88.00 &            &          &  {\bf (1)}   \\ 
 137 & J1927+1725  & J1927+1725  &  J1928+17   &  0.29 & 136.00 &  1.26e+07  & 3.16e+11 &  {\bf (1-P)} \\ 
 138 & J1928+15    & J1928+15    &             &  0.40 & 242.00 &            &          &  {\bf (1)}   \\
 139 & J1929+11    & J1929+11    &             &  3.22 &  80.00 &            &          &  {\bf (1-P)} \\
 140 & J1930+0104  & J1930+0104  &             &       &  42.00 &            &          &  {\bf (6)}   \\
 141 & J1931+42    & J1931+42    &             &       &  50.90 &            &          &  {\bf (7)}   \\
 142 & J1944-10    & J1944-10    &             &  0.41 &  31.01 &            &          &  {\bf (1)}   \\
 143 & J1946+24    & J1946+24    &             &  4.73 &  96.00 &            &          &  {\bf (1)}   \\
 144 & J1952+30    & J1952+30    &             &  1.67 & 188.60 &            &          &  {\bf (1)}   \\
 145 & J1956-28    & J1956-28    &             &  0.26 &  45.69 &            &          &  {\bf (1)}   \\
 146 & J1958+30    & J1958+30    &             &  1.10 & 199.30 &            &          &  {\bf (1)}   \\
 147 & J2000+29    & J2000+29    &             &  3.07 & 132.50 &            &          &  {\bf (1)}   \\
 148 & J2007+20    & J2007+20    &             &  4.63 &  67.00 &            &          &  {\bf (1)}   \\
 149 & J2033+0042  & J2033+0042  &  J2033+00   &  5.01 &  37.84 &  8.20e+06  & 7.05e+12 &  {\bf (1)}   \\
 150 & J2047+1259  & J2047+1259  &             &       &  36.00 &            &          &  {\bf (8)}   \\ \bottomrule
\end{tabular}
\end{minipage}
\label{t_rrat3}
\end{table*}
\begin{table*} 
\caption{List of RRATs - continued.}
\label{t_list04}
\begin{minipage}{0.95\textwidth}
\centering
\begin{tabular}{|l|l|l|l|l|l|l|l|l|} \toprule
     &  B-Name     & J-Name      & Other-Name  & P$_s$ &  DM    &  $\tau$    & B$_s$    &        \\
     &             &             &             &  s    & pc.cm$^{-3}$ & yr    & G        &        \\ \midrule
 151 & J2052+1308  & J2052+1308  &             &       &  42.00 &            &          &  {\bf (6)}   \\ 
 152 & J2105+19    & J2105+19    & J2105+1917  &       &  33.00 &            &          &  {\bf (6)}   \\
 153 & J2105+6223  & J2105+6223  &             &  2.30 &  50.75 &  7.00e+06  & 3.51e+12 &  {\bf (1)}   \\
 154 & J2107+2606  & J2107+2606  &             &       &  10.50 &            &          &  {\bf (6)}   \\
 155 & J2135+3032  & J2135+3032  &             &       &  63.00 &            &          &  {\bf (6)}   \\
 156 & J2146+2148  & J2146+2148  &             &       &  43.00 &            &          &  {\bf (6)}   \\
 157 & J2202+21    & J2202+21    & J2202+2147  &       &  17.00 &            &          &  {\bf (6)}   \\
 158 & J2205+2244  & J2205+2244  &             &       &  22.00 &            &          &  {\bf (6)}   \\
 159 & J2210+2118  & J2210+2118  &             &       &  45.00 &            &          &  {\bf (6)}   \\
 160 & J2225+35    & J2225+35    &             &  0.94 &  51.80 &            &          &  {\bf (1)}   \\
 161 & J2311+67    & J2311+67    &             &  1.94 &  97.10 &            &          &  {\bf (1-P)} \\
 162 & J2325-0530  & J2325-0530  &             &  0.87 &  14.97 &  1.34e+07  & 9.57e+11 &  {\bf (1)}   \\ \bottomrule
\end{tabular}
\end{minipage}
\label{t_rrat4}
\end{table*}

\end{document}